\documentclass{cernyrep}
\usepackage{texnames}
\usepackage[T1]{fontenc}
\usepackage{float}

\usepackage{fancyhdr}
\fancyhfoffset{4 mm}
\fancypagestyle{ARTTITLE}{%
\fancyhf{} 
\lhead{\small{CERN Accelerator School Proceedings ---
{\it Advanced Accelerator Physics} ---  Spa,  Belgium, 2024}}
\lfoot{\hspace{3mm} Available online at \url{https://cas.web.cern.ch/previous-schools}}
\rfoot{\thepage\hspace*{3mm}}
 
}

\usepackage{subfig}
\usepackage{amsmath}

\usepackage{listings}
\usepackage[labelfont=bf]{caption}
\usepackage{varwidth}
\usepackage{xcolor}
\usepackage[breaklinks=true, bookmarks, colorlinks=true, linktoc=page, pdftex, linkcolor=black, citecolor=black, urlcolor=blue]{hyperref}
\begin{document}
\title{LHC operation and the High-Luminosity LHC upgrade project}
\author{O. Br\"uning and M. Zerlauth}
\institute{CERN, Geneva, Switzerland}


\begin{abstract}
This paper describes the main concepts and performance goals for the LHC and HL-LHC projects. It summarizes the main technical challenges and highlights the key technologies that have been developed for both projects. 
\end{abstract}

\maketitle
\thispagestyle{ARTTITLE}
\section{Introduction}

The performance of a hadron collider can be measured by 3 key performance parameters:
\begin{itemize}
\item \textbf{The Center of Mass [CM] collision energy:} The Standard Model without the Higgs particle provides contradictory predictions for CM collision energies above 1 TeV [also referred to as the~unitarity problem of the standard model without Higgs particle]. The LHC was therefore designed as a Higgs discovery machine with sufficient headroom for the CM collision energy to cover a wide range of possible Higgs masses. The Higss particle was finally successfully discovered in the LHC in 2012 at a surprizing low mass of 125.35 GeV. A collider providing CM energies in excess of 251 GeV is therefore able to produce the Higgs particle and to study its properties. Choosing protons as the colliding particles for the LHC and HL-LHC projects implies required beam energies in excess of 251 GeV for the Higgs production. Protons are not fundamental particles and consist themselves of quarks and gluons. The collisions in the LHC are therefore actually collisions of quarks and gluons, each of which is carrying only a fraction of the total proton momentum. As a~proton consists of 3 quarks and gluons, the Higgs particle could already be discovered with beam energies of 3.5~TeV used in the first operational run of the LHC. Higher CM collision energies are still desirable in order to look for physics Beyond the Standard Model [BSM] and efforts continue in the LHC to increase the beam energy towards the nominal design beam energy of 7~TeV. The~initial operation periods of the LHC were limited to beam energies of 3.5 TeV for the first year of operation and 4~TeV for the remainder of the first running period. The second running period increased the beam energy to 6.5 TeV and the third running period to 6.8 TeV. The HL-LHC upgrade still aims for operation at the nominal beam energy of 7 TeV.
\item \textbf{Instantaneous luminosity:} The instantaneous luminosity is defined as the rate at which a collider produces events in a detector. The measured rate in the detector is given by the product of the instantaneous luminosity and the cross section of the event of interest. Looking for rare events (such as the Higgs discovery), a minimum luminosity of $10^{33} cm^{-2}s^{-1}$ was assumed to be necessary for the Higgs discovery. The instantaneous luminosity is entirely determined by the accelerator and beam parameters and proton particles were chosen as the best candidate for achieving such high luminosity levels and covering a broader energy range in the collisions for the discovery of particles with yet to be determined characteristics.
\item \textbf{Integrated luminosity:} In the end, the success of the experiments relies on a statistical analysis of data and the performance discovery reach of the detectors relies to a large extend on the total sample of measured events. Therefore, what matters most for the experiments is less the instantaneous luminosity but rather the integral of the produced luminosity over time. This quantity describes then the total data volume accessible to the experiments and depends in addition to the~machine and beam parameters on the overall machine efficiency and availability and the total time scheduled to produce data – i.e. the physics run period. The integrated luminosity is often measured in inverse barns, where 1 barn is $10^{-28}$ ${m^2}$. The goal for the LHC machine was the production of the order of 300 $fb^{-1}$. The worldwide data produced in hadronic collisions prior to the LHC amounted to approximately 11 $fb^{-1}$, mainly coming from the operation of the Tevatron collider at Fermilab, underlining the ambitious goal of the LHC collider project. The HL-LHC upgrade aims at a further 10 fold increase in the nominal integrated luminosity to about 3000 $fb^{-1}$.
\end{itemize}

\begin{figure}[ht]
\begin{center}
\includegraphics[width=16cm]{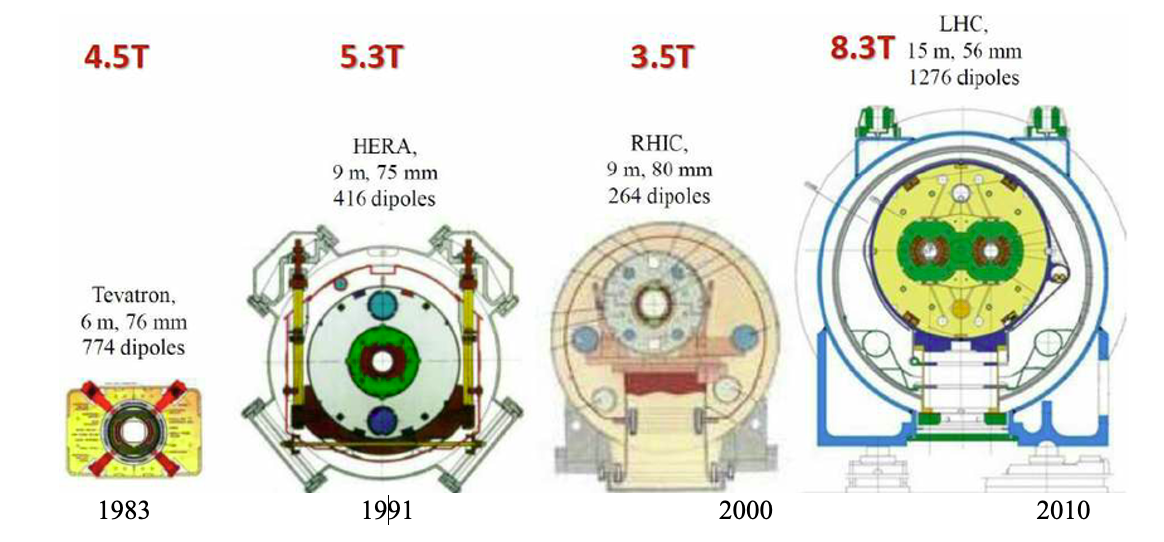}
\caption{Main dipole magnets from the Tevatron, HERA, RHIC and LHC projects.}
\label{fig:Magnets}
\end{center}
\end{figure}

\section{LHC in the LEP tunnel}
First studies of the LHC project started already in 1983, shortly after the approval of the Large Electron Positron [LEP] collider at CERN and preceding the publication of the LEP design report in 1984. The~LEP project featured a 27 km long underground tunnel in the Geneva basin, with approximately 22 km of arcs and 5 km of straight sections distributed over 8 insertions. The LEP tunnel is located approximately 100 meters underground, maximizing its extent through the molasse by featuring a 1.4 \% slope and minimizing its extent under the Jura limestone while maximizing the overall circumference. The~LHC was from the beginning conceived as a follow-up project to the LEP collider using the same tunnel infrastructure. Aiming for a proton beam energy of 7 TeV in a tunnel with 22 km of arcs, requires a magnetic bending field in excess of 8 T, clearly exceeding the performance of superconducting magnets used for previous hadron colliders and the existing NbTi based superconducting magnet technology. The~Tevatron collider at Fermilab featured dipole magnets with a peak field of 4.5 T, HERA at DESY magnets with 5.3 T and the RHIC collider at BNL magnets with 3.5 T. Pushing the magnetic field beyond 8 T therefore clearly pushed the existing NbTi superconductor technology to its limits, requiring an unprecedented operating temperature of 1.9 K to provide the required operational margins. Using superfluid He at 1.9 K did not only provide additional margins for the peak field generation with NbTi cables, but also allowed an efficient cooling of the cables as superfluid He can easily penetrate all insulation layers and very efficiently transport any heat that is deposited near the coils away from the superconductor. Featuring two counter circulating beams of the same charge requires opposite magnetic field directions in the~apertures of the two beams. The LHC magnets therefore feature a novel 2-in-1 design concept, where the~two separate magnetic coils share the same cryogenic and magnetic infrastructure. Figure~\ref{fig:Magnets} shows the~historical evolution of the NbTi superconducting dipole magnet designs.

\begin{figure}[ht]
\begin{center}
\includegraphics[width=15cm]{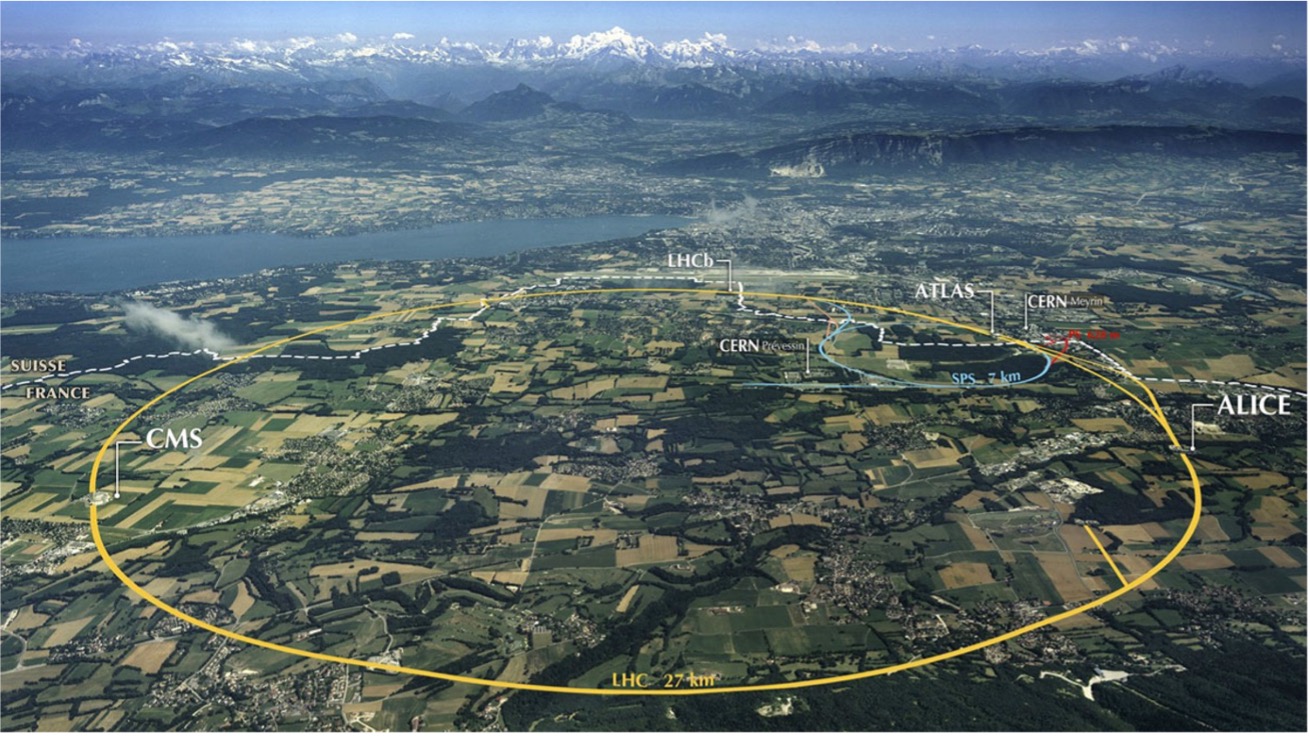}
\caption{Overview of the LHC machine layout and its 4 main experiments in the Geneva basin.}
\label{fig:LHC-Layout}
\end{center}
\end{figure}

The LHC features 1232 main dipole magnets and in total over 9000 magnetic elements, operating at 1.9 K, cooled by 150 tonnes of He [out of which 90 tonnes at 1.9 K] and powered by over 1700 power converters and protected by over 7000 quench detection systems. In addition, around 140 normal conducting magnets are installed in regions of the tunnel featuring increased levels of radiation, e.g. in the~collimation regions or close to the high luminosity experiments. The beams in the LHC are composed of approximately 2800 packages, called bunches, each containing approximately $10^{11}$ particles per bunch for reaching the LHC performance goals and translating to a net stored beam power of approximately 350~MJ per beam. The stored electromagnetic energy inside the LHC magnet system of the eight symmetric arcs exceeds 10 GJ. Energies of this order of magnitude combined with the voltages developing in the~magnet chain during powering failures requires the segmentation of the LHC magnet system into 8 independent sectors and special care and monitoring for the protection of accelerator equipment during operation. This includes a sophisticated three stage collimation system that removes any stray particles from the circulating beams before they could end up in the superconducting magnets or other sensitive accelerator component. The quench tolerance for energy deposition in the superconducting magnets is less than 20 $mW/cm^3$, corresponding to less than $10^{-6}$ particles of the nominal beam intensity impacting inside the cold-mass. In addition to operation with proton beams, the LHC operates also with $Pb^{82+}$ fully stripped ions and features 4 separate main experiments (Fig.~\ref{fig:LHC-Layout}): ATLAS and CMS as multi-purpose detectors to look for the Higgs particle, the LHCb experiment for high precision measurements of rare events to probe the accuracy of the Standard Model and to look for physics Beyond the Standard Model and the ALICE detector for in-depth studies of ion collisions.

\section{LHC Operation and Performance}

\begin{figure}[ht]
\begin{center}
\includegraphics[width=16cm]{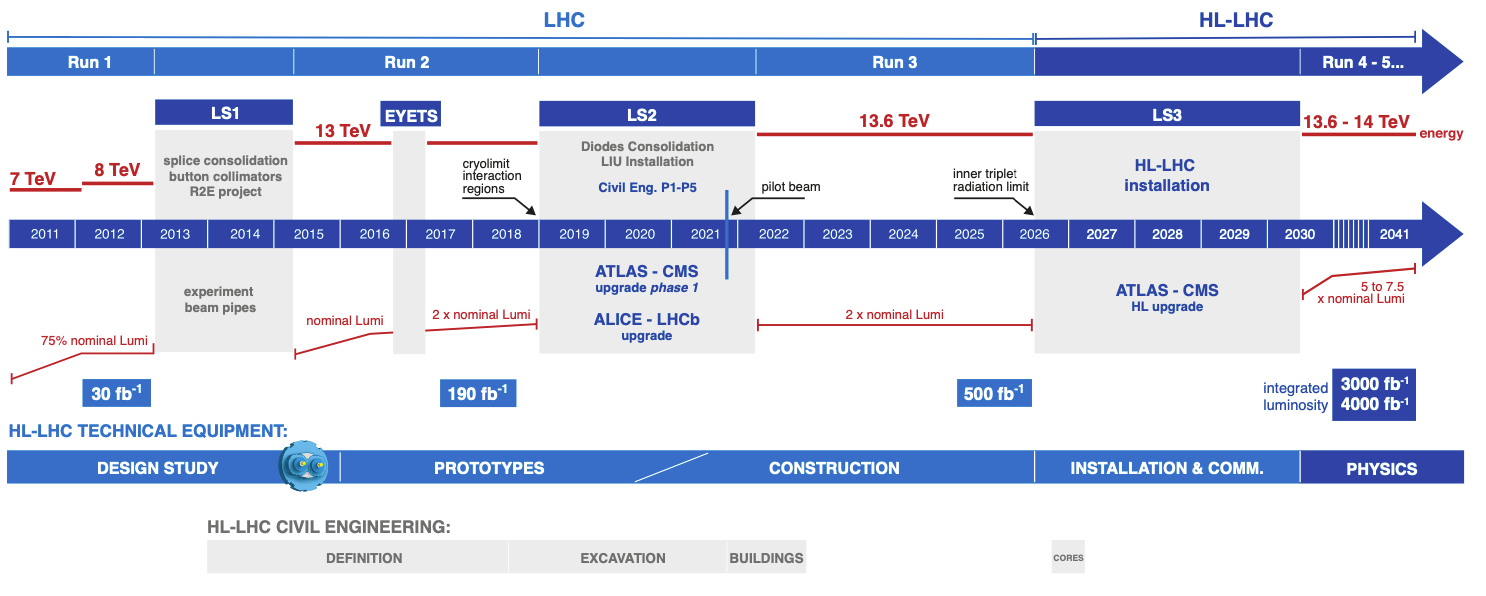}
\caption{The LHC schedule for the initial 3 running periods and the outlook to the HL-LHC era.}
\label{fig:HL-LHC-schedule}
\end{center}
\end{figure}

LEP operation was stopped in 2000, followed by the dismantling of the LEP machine and the installation work for the LHC. LHC beam operation started in 2008, but was interrupted shortly after due to a~fault of one of the magnet interconnections during the final steps of commissioning the superconducting magnet circuits. The fault resulted in an electrical arc that opened the cryogenic lines as well as the~beam and insulation vacuum systems and affected most of the magnets in sector 34 of the LHC machine. The~incident illustrates the large damage potential originating from the electromagnetic energy stored in the~magnet system. Repair of the affected area and accelerator elements took a little over one year and hardware commissioning resumed by the end of 2009 and beam physics operation restarted in 2010. The cause of the inter-magnet connection fault was a non-conform joint soldering between two adjacent magnets. As similar imperfections could not be excluded in other interconnections, the beam energy of LHC operation was initially limited to 3.5 TeV per beam for the first year of operation and then raised to 4 TeV after more measurements and analysis became available following the first year of operation. Operation at these lower beam energies were judged safe, as similar events as in 2008 could not occur at these energies and the initial operation phase was used to systematically validate all inter-magnet connections in the machine. This exercise identified several other magnet connections that had similar weaknesses, underlining that a systematic repair work was required before further increasing the beam energy towards its nominal value of 7 TeV.
After the first 3 years of operation at lower beam energies, the LHC entered its first Long Shutdown, LS1, from 2013 to the beginning of 2015 where all splices got repaired and consolidated with additional copper bypasses. Operation resumed in 2015 with a new hardware commissioning exercise where the magnet system got trained towards higher beam energies (see Fig.~\ref{fig:HL-LHC-schedule}).

\subsection{Dipole magnet training}
\label{sec:Magnet-Training}

\begin{figure}[ht]
\begin{center}
\includegraphics[width=15cm]{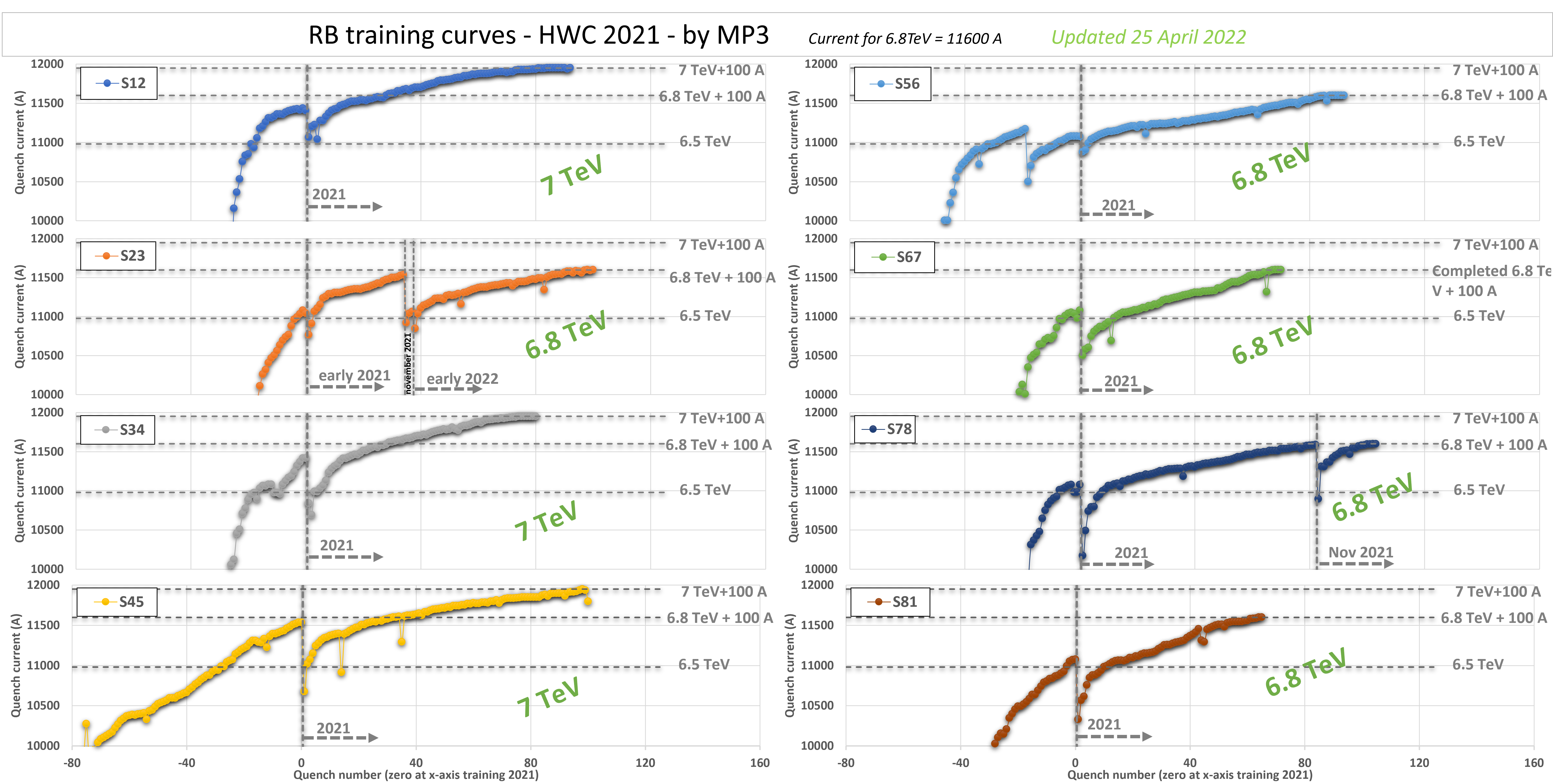}
\caption{Training history of the 8 dipole circuits of the LHC.}
\label{fig:RBTraining}
\end{center}
\end{figure}

During these initial commissioning periods, the LHC revealed another challenge on the way to higher beam energies: the system of magnets, 154 dipole magnets are powered in series in each of the eight sectors, showed a significant number of retraining quenches in the tunnel [a quench is the spontaneous loss of the superconducting state of the magnet – this can be triggered by energy deposition inside the~magnet either from beam losses or by mechanical movements of the coil windings] resulting in much longer training times than anticipated. Each magnet had been individually tested and trained to nominal operating fields before installation in the~tunnel and it was assumed that the magnets would mostly keep this ‘memory’, with only a few magnets requiring retraining once operated as a string in the~tunnel. Experience in the~tunnel however revealed a very different behavior, with some sectors showing a particularly slow training of the magnet chain towards the goal of 7 TeV. In particular, the considerable loss of memory after thermal cycles of the sectors [which are necessary during Long Shutdowns and/or for repairs on the superconducting elements] came as a surprise (see Fig.~\ref{fig:RBTraining}). This was in particular visible during the~training of the dipole chain in sector 23, which after a very fast training to 6.5 TeV during the~commissioning period of 2015, showed a considerably increased rate of retraining quenches during the~2021 and in particular the 2022 campaign, requiring an additional 54 quench events (and a~total of 64 individual magnet training quenches) to reach the operational target of 6.8 TeV.

It is important to note however that – despite the overall large number of magnet quenches in the~LHC sectors – still 45\% of all dipole magnets installed in the LHC have never experienced a re-training quench in the tunnel yet, and only 3 magnets have experienced already a maximum of 5 individual training quenches. The required operating currents can still be reached for all magnet circuits (with only a~few exceptions) 14 years after the start of the LHC, despite several thermal cycles, numerous current cycles, radiation, and a large number of training and beam induced quenches. However, every quench is a very violent process (especially in the high-current circuits) that implies a certain unavoidable risk of damage to occur (short-to-ground, internal shorts, quench heater failures, etc). The decision of the collision energy for future runs and the HL-LHC era will therefore inevitably involve a cost/benefit analysis of the required re-training effort (which implies both technical risks and the allocation of considerable time).
An example that illustrates well the risk of magnet quenches is the discovery of a~problem with the~design of the protection diode of the LHC dipole magnets after the first Long Shutdown, where a short between the diode and its metallic enclosure could appear due to the accumulation of metallic debris due to helium circulation after e.g. a magnet quench. This problem limited the number of quenches that the magnets should be exposed to and thus, the beam energy during the second operation period. The~Run~2 period therefore operated at a beam energy of 6.5 TeV. After another 4 years of operation, the~LHC machine entered another Long Shutdown, LS2, from 2019 until the beginning of 2022 where all diode boxes of the LHC magnets got refurbished, allowing in principle the operation at the nominal beam energy of 7 TeV. As the recovery time from a magnet quench in the tunnel can be quite time consuming (typically taking between 8-12 hours), the number of permissible training quenches is ultimately limited by the time that can be allocated to the magnet training. In order to keep the training time at an acceptable level, it was decided to limit the beam energy below nominal for the start of the third operation period – and thus reduce the required number of training quenches. The Run~3 operation period that started in 2022 therefore features a beam energy 6.8 TeV, still slightly short of the nominal design value. Run~3 is scheduled to be completed by the mid of 2026.

\subsection{Machine Availability}
\label{sec:Availability}
The increase in machine performance achieved during the two initial operational periods is partly thanks to an increased understanding and mastering of the machine which was demonstrated by reaching almost twice the design peak luminosity at the end of Run~2, together with an outstanding machine availability.  Large-scale research infrastructures and in particular circular colliders such as the LHC represent a major challenge in terms of equipment reliability, as many ten thousand accelerator and infrastructure components must operate simultaneously and continuously for many hours to produce the desired physics output. For the HL-LHC, the nominal fill length (including the leveling period) will be in the~order of $7.5$~hours, which will be interleaved with a turn-around time of at least $2.5$~hours to bring back the~beams into collisions after a machine failure or a deliberate termination of the prior physics fill.

Therefore, in addition to the accurate and reliable control of proton and ion beams with twice, respectively five times the stored beam energy of the nominal LHC design, machine operation in the~HL-LHC era will also require further improvements of the already outstanding machine availability that was steadily improved during the first two operational runs. 
\begin{figure}[ht]
\begin{center}
\includegraphics[width=15cm]{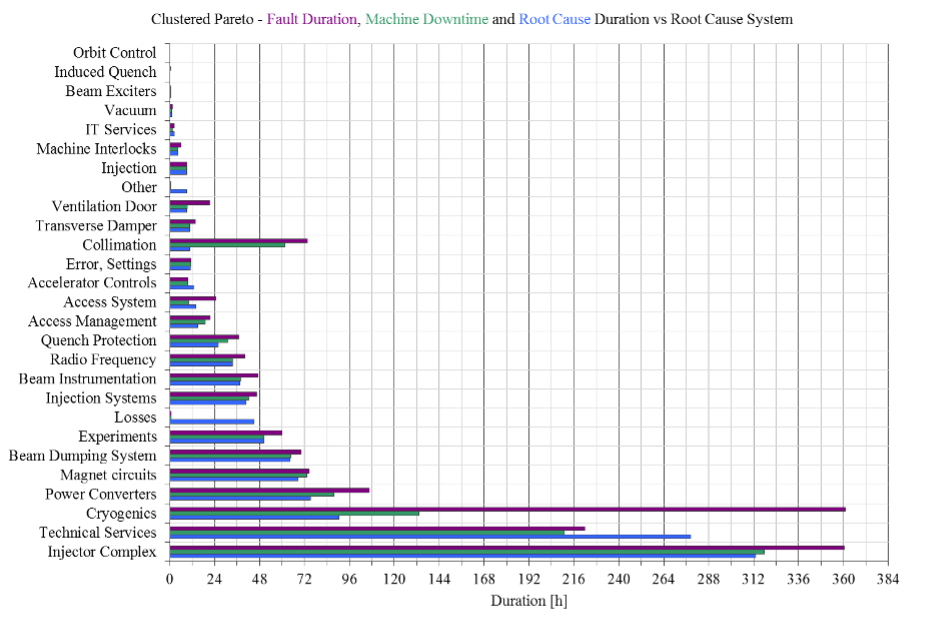}
\caption{Root causes and their integrated duration of downtime of the LHC machine in Run~2.}
\label{fig:Availability}
\end{center}
\end{figure}

During the three final years of its second operational run, the LHC managed to produce particle collisions during almost half the time devoted to high-intensity proton operation, while the remaining time was equally shared between equipment failures and regular operational time (such as the injection of beams from the injectors or the energy ramps).
This is an unprecedented achievement for such a~complex machine, which in addition is using many novel technologies that were never used at such large industrial scales before. One of the main reasons for this achievement is that dependability considerations were a~fundamental part of every equipment design from the very beginning. This is in particular the case for the backbone of the machine protection system, for which state of the art reliability engineering methods were employed to guarantee meeting both, the challenging reliability as well as availability targets.
The second, equally important ingredient is a continuous identification and documentation of the root causes of down-time arising during the operational periods of the accelerator equipment. A~dedicated tool, the~so-called Accelerator Fault Tracker (AFT) has been developed to this end, allowing to identify and quantify the impact of recurring equipment failures on machine operation and trigger targeted consolidation activities to mitigate these weaknesses.

Examples are major consolidation or displacement activities for electronics installed close to the~accelerator which have shown weaknesses to radiation induced effects, the optimization of interlock levels across numerous protection systems based on beam operation experience and the preventive replacement of several thousands of local power supplies with unsatisfactory reliability. Another important ingredient is the development of more and more powerful operational tools, both for the diagnostic of the machine state as well as for the automation of recurrent operations and adjustments, which ensure repeatability while avoiding as much as possible human errors in the execution of the complex operational sequences. 

Following these continued efforts, LHC availability has today reached a level where it is dominated on one hand by the availability of its injector complex, and on the other hand by a few, but often long stops in the infrastructure systems necessary for the operation of the large superconducting magnet system as shown in Fig.~\ref{fig:Availability}. The full injector complex underwent a major upgrade program during LS2 \cite {LIU}, the impact of which on overall availability will only become visible during the upcoming Run~3. Failures in particular in the LHC cryogenic system on the other hand will, despite often minor root causes, require many hours to recover nominal operating conditions as employing redundancy techniques is only possible to a very limited extent in such large-scale industrial systems. 
As the complexity of the LHC will further increase with the deployment of the HL-LHC upgrade, it is therefore important to maintain and even further improve the availability for HL-LHC operation. This is not only true for the newly installed machine components, but also for the remaining parts of the machine which are based on components that will approach the end-of-life at the time of the HL-LHC era. Pursuing preventive maintenance and consolidation activities as well as further improvements of intervention procedures are therefore a necessity, limiting as much as possible the need of physical access to the tunnel to perform corrective actions.

\subsection{UFOs}
\label{sec:UFOs}

Controlling the losses of highly energetic particles in a superconducting accelerator is a continuous and challenging task, especially for localized loss events which can be caused by fast beam instabilities or interactions of the proton beams with dust particles (UFOs). The latter were the main reason of beam-induced quenches in Run~2 and Run~3 and they are expected to remain the primary source of transient beam loss events in future runs. When entering the beam, dust particles get rapidly ionized and are repelled from the circulating protons within a few turns of the beams. While most events are harmless, a~small fraction of dust particles can still induce sufficient beam losses to perturb beam operation. This fraction will however increase in future runs and in particular the HL-LHC era due to the more challenging operational conditions.

\begin{figure}[ht]
\begin{center}
\includegraphics[width=15cm]{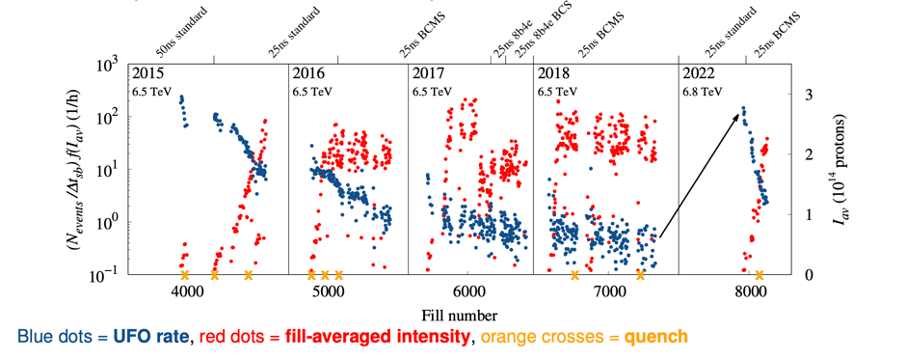}
\caption{UFO rates vs beam intensities and quench events during LHC Run~2.}
\label{fig:UFORates}
\end{center}
\end{figure}

An efficient protection of the magnets against beam-induced quenches requires an in-depth understanding of the underlying physics of the energy deposition mechanisms as well as the quench limits of the different superconducting magnets. Both have been extensively simulated and empirically studied during the first two operational runs of the LHC, allowing to define a good compromise between beam loss protection settings and beam-induced magnet quenches. Figure~\ref{fig:UFORates} shows the UFO rates observed at $6.5$~TeV in Run~2 (blue dots) as well as the beam intensity (red dots). Events which resulted in a quench of a dipole magnet are displayed as yellow crosses. A possible future increase of the beam energy to 7 TeV reduces the quench limit of the main dipole magnets by approximately $20$~\%, while the same number of lost protons will lead to $7-8$~\% higher energy densities in the magnet coils. This increased likelihood of beam induced quenches and in general reduced operational margins will require further optimizations of the protection thresholds and strategies as a function of the experience gained when operating at increased beam energies approaching the nominal energy of $7$~TeV.

\subsection{Heat Load and electron cloud}
\label{sec:Ecloud}
The LHC and HL-LHC cryogenic magnets are equipped with actively cooled beam screens, which intercept beam induced heating mainly due to synchrotron radiation, impedance and e-cloud effects.
During the LHC Run~2, large heat loads were observed on the beam screens during operation with the nominal bunch spacing of 25 ns. In particular the heat loads in some of the arcs reached levels close to the design cooling capacity of~{8}\,kW/arc. In all sectors, the heat-loads were significantly larger than expected from impedance and synchrotron radiation.

By analyzing the heat load data collected during Run\,2 and comparing them against models and simulations, it was possible to conclude that a dominant fraction of the observed heat loads is due to electron cloud effects, as a result of a larger than expected Secondary Electron Yield of the beam screen surfaces. During the Long Shutdown 2 (2019-2022) surface analyses were conducted of beam screens extracted from the accelerator, which identified specific surface modifications associated with the magnets showing the highest heat load, namely the presence of cupric oxide (CuO) and a very low carbon concentration. These modification are associated a larger Secondary Electron Yield (SEY) and therefore with a stronger e-cloud.

\begin{figure}[ht]
\begin{center}
\includegraphics[width=15cm]{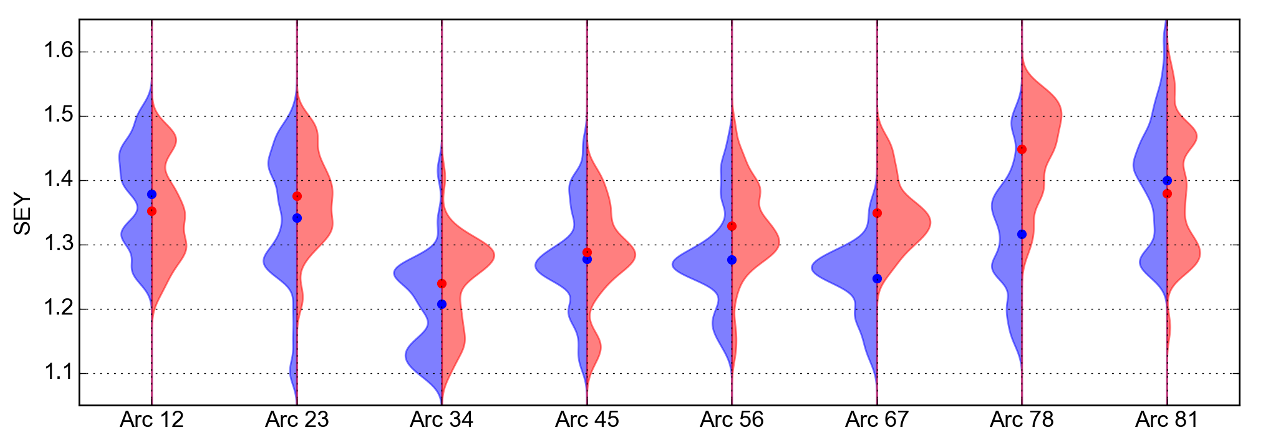}
\caption{Evolution of Secondary Electron Yield in the 8 LHC sectors between 2018 (purple on left) and 2022 (red on right).}
\label{fig:SEYarcs}
\end{center}
\end{figure}

Numerical simulations can be used to estimate the arc heat loads expected for the HL-LHC beam parameters. Figure\,\ref{fig:SEYarcs} shows the change of Secondary Electron Yield in the 8 LHC sectors between 2018 and 2022, showing a considerable degradation following the venting of the sectors during the second long shutdown (e.g. sector 67 and sector 78), while for other sectors the situation remained stable or even improved (e.g. sector 12).

During Run~2 and Run~3, the LHC cryogenics has been operated in an optimized configuration (using one cold-compressor unit to serve two consecutive sectors). The cryoplants feeding the high-load sectors have been recently characterized by the cryogenics team, and they were found to perform better than their design specifications, being able to deliver 10 kW/arc.
Despite this increased cooling capacity, the additional degradation of the SEY observed after LS2 is not compatible with the HL-LHC nominal beam configuration. Surface treatments will therefore become necessary during LS3 in order to reduce the SEY of the beam-screen surfaces. Alternatively hybrid filling patterns are already being used during recent LHC operational runs to partially mitigate the performance loss.

\subsection{Performance}
\label{sec:Performance}
\begin{figure}[ht]
\begin{center}
\includegraphics[width=12cm]{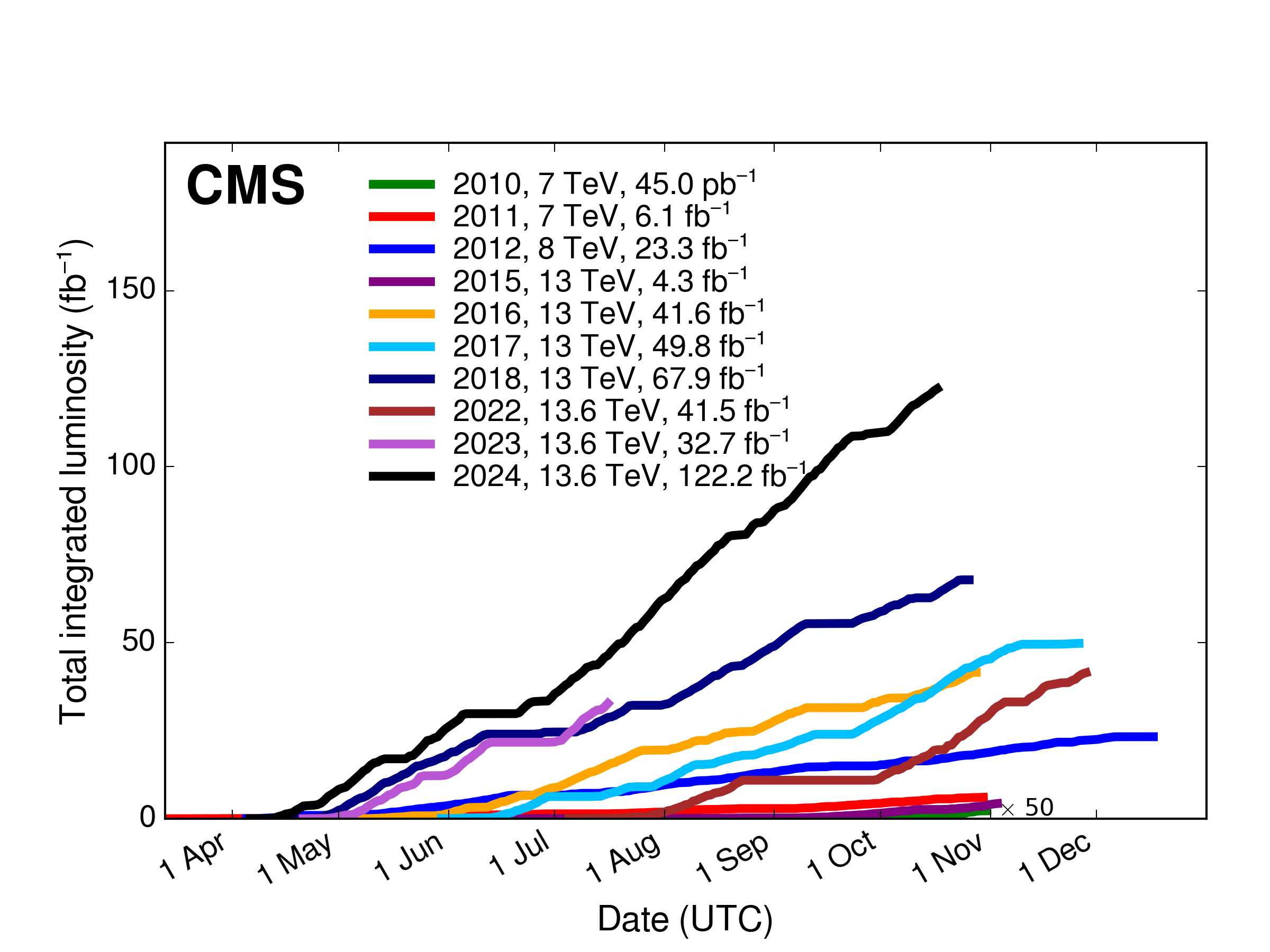}
\caption{The Integrated Luminosity recorded by the CMS detector for the different operational years.}
\label{fig:CMSLumi}
\end{center}
\end{figure}

In spite of the limitation of the beam energy, LHC operation showed otherwise a remarkable performance during the first three operational runs! The luminosity increased continuously with every operational year and quickly surpassed the nominal design value. Also, the machine availability and efficiency of the LHC were remarkable already in the early years of operation, allowing the integrated luminosity to surpass the yearly targets and to allow the Higgs discovery already during the first running period in 2012 \cite {Higgs}. Figure~\ref{fig:CMSLumi} shows the achieved integrated luminosity over the first operational periods of the~LHC, up to the~third year of Run~3. The performance reached about 160 $fb^{-1}$ during the first two operation periods, with an annual integrated luminosity of about 70 $fb^{-1}$ towards the end of the Run~2, paving the way for a total integrated luminosity of about 500 $fb^{-1}$, well above the original design goal of 300 $fb^{-1}$, by the end of the LHC operation period in 2026. The integrated luminosities 2022 and 2023 fell slightly short of the~projected $70fb^{-1}$. In 2022 the luminosity production suffered from a~long interruption due to problems with the superconducting RF system and in 2023 the proton run was prematurely terminated due to vacuum problems in the triplet region of Point 8. However, during both years, the slope of the~integrated luminosity production versus time was equal or even steeper when compared to the year with the~highest integrated luminosity in 2018. The latest operational year of the LHC in 2024 on the other hand was truly spectacular, allowing to accumulate $122.2fb^{-1}$ in the high-luminosity experiments of ATLAS and CMS. While the beam parameters were mostly unchanged with respect to the previous year 2023, this performance was made possible by a very high availability of the LHC, without any major faults or interruptions all along the year.

\begin{figure}[ht]
\begin{center}
\includegraphics[width=15cm]{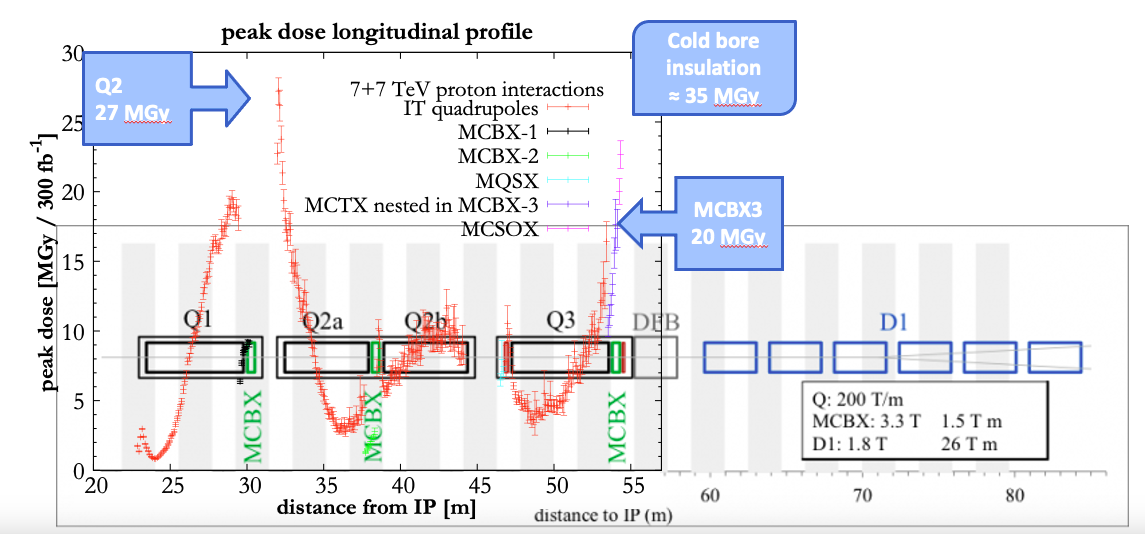}
\caption{Triplet layout and expected radiation dose in the ATLAS and CMS insertion regions for an~integrated luminosity of 300 $fb^{-1}$. }
\label{fig:RadDamageTriplet}
\end{center}
\end{figure}

For integrated luminosities above 300 $fb^{-1}$, it is expected that the radiation damage in the focusing quadrupole magnets next to the high luminosity experiments of ATLAS and CMS, the so called triplet quadrupoles as they are composed of three separate units, will start to compromise the performance of the magnet system. Figure~\ref{fig:RadDamageTriplet} shows the schematic layout of the triplet magnets, with the experiment being on the left side of the figure. Overlaid to the layout is the expected radiation dose, assuming a constant operation configuration throughout the LHC operation. The simulations show that the radiation dose inside the magnets largely exceeds at some places the value of 10 MGy, a radiation level where most epoxy and insulation materials become brittle and lose their mechanical integrity. Therefore, the triplet magnets must be replaced by the end of the LHC Run3 period when it is projected to have accumulated over $500 fb^{-1}$ if the operation is to continue until June 2026 \cite {AUP} \cite {WP3}.
This exceeds by a large margin the initial design luminosity of the LHC $300 fb^{-1}$ and as such the estimated lifetime of the present LHC triplet magnets, This is however believed to be possible thanks to a change/ variation of the optics polarity and crossing angle polarity so that the radiation exposure will be spread more uniformly across the triplet magnets, enhancing their overall lifetime.

\section{The HL-LHC project}
\label{sec:HL-LHC}
The HL-LHC upgrade project aims at extending the LHC machine lifetime well beyond the nominal running period, so that the LHC can be operated until the early 2040ies, and devise beam parameters and a machine configuration that allows collecting 3000 $fb^{-1}$ of data over the HL-LHC operation period, a~tenfold increase over the~nominal LHC design goal. Achieving this goal implies the production of ca. 250 $fb^{-1}$ per year, or in other words, the production of almost the total nominal LHC design luminosity within a given year of HL-LHC operation. A conventional approach would be to increase the instantaneous luminosity proportionally to this goal. However, the LHC experiments presented another design criterion for the HL-LHC that specified that not too many collisions should be produced for each bunch crossing in order to assure an efficient detector operation and data analysis by the experiments. The experiments therefore requested that the instantaneous luminosity should not be larger than 5 times the~nominal LHC design goal: 5.0~$10^{34} cm^{-2}s^{-1}$  compared to the~nominal LHC value of 1.0~$10^{34} cm^{-2}s^{-1}$. In this context one should underline that the LHC operation already surpassed this design goal and achieved instantaneous luminosity of 2.0 $10^{34} cm^{-2}s^{-1}$. Achieving the ambitious HL-LHC design goal of 3000~$fb^{-1}$ therefore requires for the HL-LHC in addition to the increase in the instantaneous luminosity, an increase in the overall machine efficiency and availability and the implementation of a novel leveling operation mode. For this, the machine is designed for a higher instantaneous luminosity than is digestible by the experiments, and where operation maintains this value in a controlled manner at a~level that is compatible with the detector operation during physics production. This concept requires new operational tools as well as a high machine reliability as any premature termination of a fill will imply an~even more substantial loss of integrated luminosity compared to today's LHC operation.

\subsection{Luminosity Levelling}
\label{sec:LumiLevelling}
Different options for the luminosity levelling have been explored since the start of LHC operation. One of the first tools tested and implemented in the LHC operation was the levelling via transverse beam offsets. Initially, this scheme was requested by the ALICE experiment for data taking during the proton beam operation. Following initial tests, this method has been fully validated and implemented as a tool for all 4 LHC experiments. Another levelling option is the levelling through the beam crossing angle at the~interaction points. As we will discuss later in the context of the Crab Cavity RF system, this would have been a very elegant tool for the luminosity levelling. However, this method does not reduce the~density of events inside the LHC detectors but effectively only changes the length over which collisions will be produced within the experiments. As the event density is ultimately the key limitation for the~detector performance, this method was judged in the end as not very appealing for the experiments. The third option for luminosity levelling is the variation of the beam size at the collision point through changes in the magnetic focusing of the beams left and right from the detector. While this method is clearly the~most elegant levelling technique, it is also the most complex technique for the operation of the machine, as changes in the focusing strength will affect also the crossing angle and beam orbit at the interaction point with potential implications on the machine protection systems. However, all three techniques have by now been successfully demonstrated in LHC operation and have been established as standard operational tools for the LHC operation.

\subsection{HL-LHC Technology developments}
\label{sec:HLTechnology}
We mentioned in section \ref{sec:Performance} already the limitation of the initial triplet magnets and the need to exchange these magnets as part of the HL-LHC upgrade. In addition to this central upgrade requirement, the HL-LHC upgrade drives additional vital and cutting-edge developments in other technological areas. In the following, we list a brief summary of the new key technologies that are developed for the HL-LHC upgrade.    

\subsubsection{New Triplet Quadrupole Magnets}
\begin{figure}[ht]
\begin{center}
\includegraphics[width=12cm]{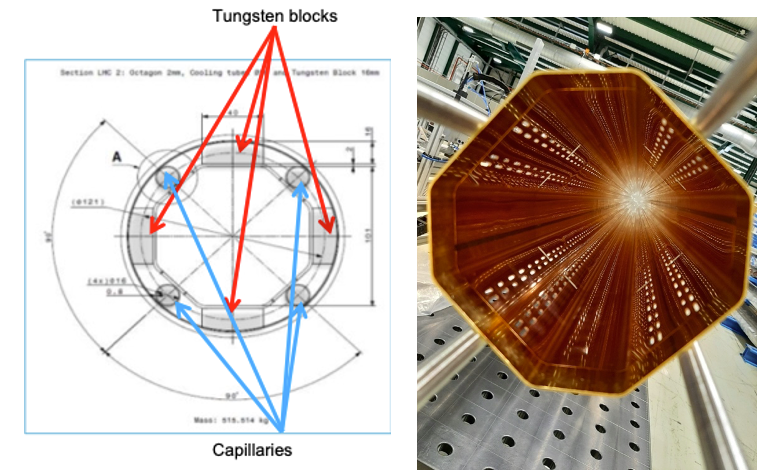}
\caption{Left, the schematic cross section of the new HL-LHC beam screen for the inner triplet magnets, on the right a first beam screen prototype.}
\label{fig:MagnetBS}
\end{center}
\end{figure}

Extending the triplet lifetime by a factor 10 from 300 $fb^{-1}$ to 3000 $fb^{-1}$ is only possible via the implementation of active shielding that protects the radiation sensitive materials in the coils from the particle fragments that escape from the interaction points. For the HL-LHC triplet magnets this is implemented by a novel octagonal beam-screen design that features ca. 2 cm thick tungsten blocks between the beam screen and the magnet cold-bore. The heat deposited inside the tungsten blocks from the beam fragments escaping the interaction points is removed by an active beam screen cooling system at temperatures around 60 K before the heat could escape into the magnet cold-mass that is operating at 1.9 K. Figure~\ref{fig:MagnetBS} shows the schematic cross section of the new HL-LHC triplet magnets with the octagonal beam screen, tungsten blocks (highlighted in red) and the active cooling capillaries (highlighted in blue).   

\begin{figure}[ht]
\begin{center}
\includegraphics[width=16cm]{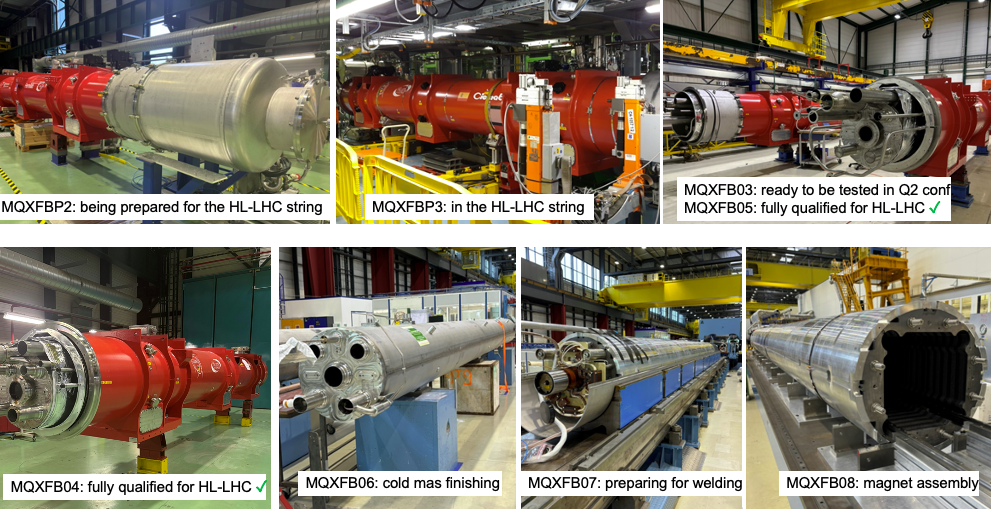}
\caption{Status of Main Quadruple Magnet (MQXFB) production at CERN.}
\label{fig:MQXFB}
\end{center}
\end{figure}

Installing the additional tungsten absorbers requires a larger magnet aperture. In addition, the~HL-LHC beam operation targets smaller beam sizes at the collision points, implying larger beam sizes and crossing angles inside the triplet magnets, which implies a further increase in the magnet aperture. The~HL-LHC upgrade therefore aims at a coil aperture of 150 mm, compared to the 70 mm coil aperture in the present LHC triplet magnets. As the triplet functionality is defined by the magnet gradient, a larger aperture ultimately implies a higher peak field inside the coils for a given magnet gradient. The nominal LHC configuration featured triplet gradients of 210 T/m, implying a peak field of about 8 T in the triplet magnet coils. Doubling the magnet aperture and keeping the gradient constant would imply peak fields of ca. 16 T at the magnet coils, a field level that is incompatible with available magnet technologies. As a compromise, the HL-LHC upgrade reduced the triplet magnet gradient to 140 T/m, implying a longer length of the triplet magnet string to achieve the required focusing strength.
Production of the new inner triplet quadrupole magnets is ongoing in full swing at CERN (long version MQXFB) and AUP (short version MQXFA), with close to half of the production being completed as can be seen in Fig.~\ref{fig:MQXFB}.

\subsubsection{New Underground Civil Engineering}
The higher luminosities in the HL-LHC era imply additional heat loads in the triplet magnets and require additional cooling capacities for the insertion magnets. Furthermore, the goal of high efficiency and highly reliable operation of the HL-LHC implies that sensitive electronic components are removed from the tunnel areas where they are exposed to ionizing radiation from proton losses and instead moved to areas where access during operation is facilitated. This considerably eases preventive maintenance and shortens the intervention times. To this end, the HL-LHC project planned for new underground structures and caverns to house the new cryogenic cold boxes and distribution equipment, the new infrastructure and powering equipment along with most of the electronic racks that had previously been installed in the LHC tunnel areas. The new underground constructions feature at each of the 2 high luminosity interaction regions a new cavern and an approximately 300-meter-long gallery and two 50-meter-long service tunnels on both sides of the experiments that connect the new gallery to the existing LHC tunnel.

\begin{figure}[ht]
\begin{center}
\includegraphics[width=16cm]{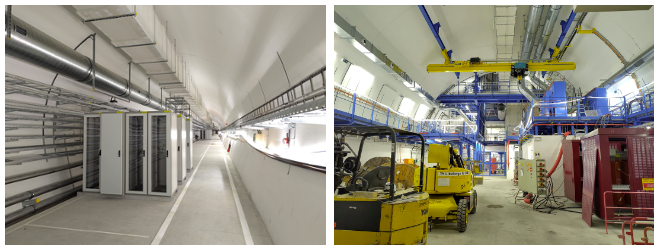}
\caption{The new underground installation. Left: View into the new underground galleries, housing the~infrastructure and powering equipment of the new long-straight sections. Right: View of the new HL-LHC underground cavern and access shafts, housing electrical distribution and cryogenic equipment.}
\label{fig:CE}
\end{center}
\end{figure}

In addition to the new underground structures, the HL-LHC upgrade features 5 additional surface buildings on the existing ATLAS and CMS sites. The new underground areas are constructed in what has been labelled the ‘Double Decker’ configuration, where the new underground areas are located approximately 10 meters above the existing LHC tunnel structure and where the connection of the new and old structures is established via 14 vertical cores of ~1 m in diameter for each of the two sites. This configuration provides optimum shielding of the new areas against radiation effects from the existing LHC tunnel and facilitates the connection of the two structures. The bulk of the civil engineering work has been conducted during LS2, when the LHC machine was not operating, in order to minimize the~perturbative effect of the heavy civil engineering work on the LHC operation. All underground civil engineering work has been successfully terminated by the end of 2022 and all new surface buildings have been handed over to CERN by January 2023. CERN is now conducting the installation of the technical infrastructures, like cooling and ventilation, electrical distribution, and emergency communications. The~installation of the~new HL-LHC equipment is already ongoing since the end of 2023. Figure~\ref{fig:CE} shows the new underground installations for the ATLAS site (with the CMS structure being almost identical).

\subsubsection{Crab Cavities}
The longer triplet magnet length and the reduced beam size at the Collision Points requires a larger crossing angle of 500 $\mu$rad compared to the 285 $\mu$rad LHC design value and the 320 $\mu$rad, currently used for LHC operation. The crossing angle mitigates the detrimental perturbations from the non-linear beam-beam interactions and unwanted [parasitic] bunch crossings along the common beam pipe between the two D1 magnets left and right from the IPs.

\begin{figure}[ht]
\begin{center}
\includegraphics[width=14cm]{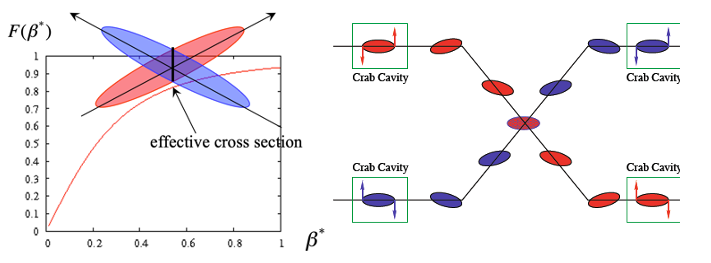}
\caption{Functioning of Crab Cavities. Left side: the luminosity reduction factor as a function of $\beta$* for a~minimum normalized beam separation of 10 $\sigma$. Right side: schematic illustration of the transverse bunch deflection that reconstitutes a~perfect beam overlap at the IPs.}
\label{fig:CCCrossing}
\end{center}
\end{figure}

Combined with a reduction of the optical $\beta$-function from 55 cm in the nominal LHC design and 30 cm in the current LHC operation period to 15 cm at the Interaction Points (IP) for the HL-LHC configuration, the overlap of the two colliding bunches at the IP is considerably reduced and thus the attainable luminosity by about 70\%. Figure~\ref{fig:CCCrossing} illustrates this effect and shows the geometric luminosity reduction factor as a function of the optical $\beta$-function at the IP for a minimum normalized beam separation of 10 $\sigma$ at the parasitic collision points. To recover the luminosity, two different Crab Cavity [CC] designs that allow for a transverse bunch rotation were developed in collaboration with the LARP collaboration: a Double Quarter Wave [DQW] design optimized for a crossing angle in the vertical plane and an RF Dipole design [RFD] optimized for a crossing angle in the horizontal plane. 

\begin{figure}[ht]
\begin{center}
\includegraphics[width=12cm]{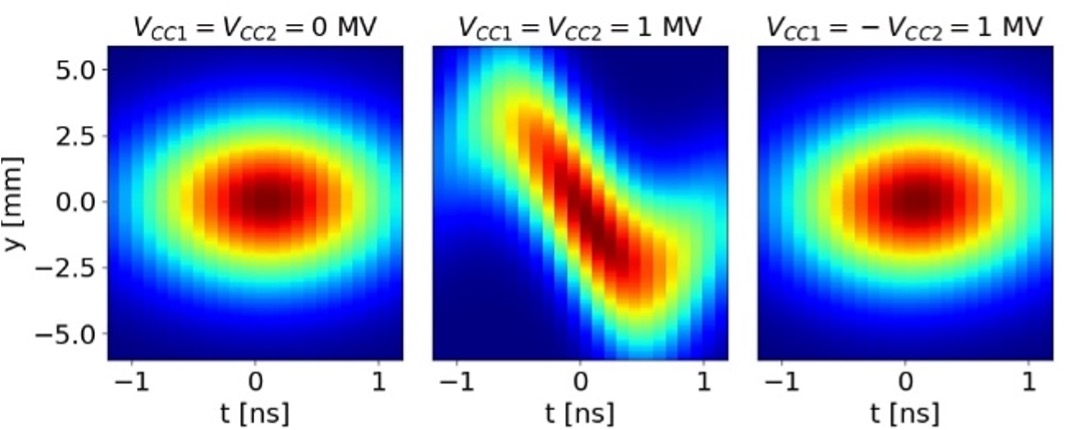}
\caption{Intrabunch motion from three different cases. Left: Cavities switched off (V = 0V). Centre: Synchronous crabbing with both cavities in phase (V = 2 MV). Right: Cavities in counter-phase, corresponding to an effective zero kick voltage (V < 60 kV).}
\label{fig:CCBeam}
\end{center}
\end{figure}

A DQW prototype experimental demonstration of crabbing of a hadron beam was realized. Several key beam experiments have been conducted in CERNs Super Proton Synchrotron (SPS) as part of a test campaign that started in 2018 with high energy proton beams of up to 450 GeV in the presence of crab cavities. Figure~\ref{fig:CCBeam} shows the measured bunch rotations in the SPS.

\subsubsection{New Absorbers and Collimators}
The higher beam intensities in the era of HL-LHC operation imply the use of more robust collimator and absorber materials that feature at the same time a lower impedance than those adopted for the~start of LHC operation. 
The increase in total beam energy in the HL-LHC era, close to 700 MJ as compared to the~approximately 350 MJ of the nominal LHC configuration, implies in addition an upgrade of all the~beam dump windows and the core of the beam dump absorber. Observations during the Run~2 operation period of the LHC, have given indications that the current beam dump core is already degrading from the~nominal LHC operation, indicating that an upgrade of the beam dump and its core material is required for reliable operation in the HL-LHC era.
The new challenges related to operation at lower $\beta$* and at higher peak luminosity also required an upgrade of the absorber blocks at the Machine – Detector interface, the~TAS and TAN absorbers and a complete re-design of the collimation systems around the~high-luminosity experiments. The HL-LHC upgrade therefore features new, low impedance secondary collimators and new, additional tertiary collimators and a more performing physics-debris collimation system. The low impedance collimators feature new designs using Molybdenum coated Molybdenum-Graphite and Copper coated Graphite jaws. About half of the LHC secondary collimators have already been replaced during Long Shutdown 2 [LS2] prior to the LHC Run~3 period that started in 2022. Given the criticality of maximising the physics time at the HL-LHC, all new collimators mount in-jaw beam position monitors (based on button pick-ups) for faster alignment and easier handling of the complex levelling gymnastics.

\begin{figure}[ht]
\begin{center}
\includegraphics[width=12cm]{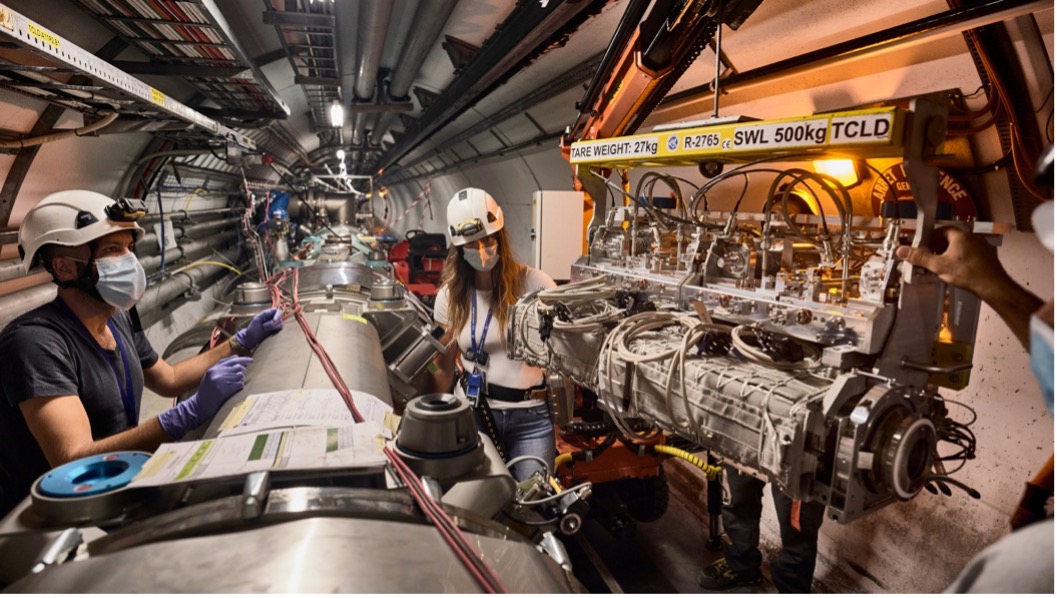}
\caption{New DS collimator assembly during installation next to ALICE experiment.}
\label{fig:DSCollimator}
\end{center}
\end{figure}

\begin{figure}[ht]
\begin{center}
\includegraphics[width=12cm]{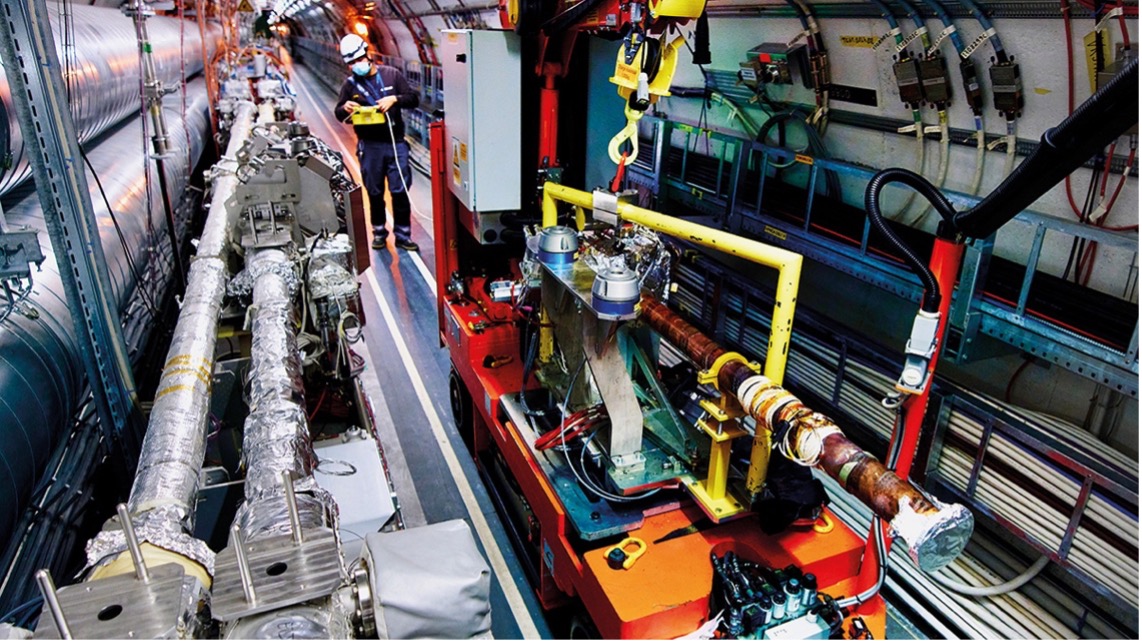}
\caption{Goniometers and the contained crystal collimators prior to their installation in the LHC.}
\label{fig:Goniometers}
\end{center}
\end{figure}

For operation with ion beams, the cleaning efficiency in the dispersion suppressors [DS] of the~ALICE experiment will be augmented by new DS collimators and by careful steering the beam trajectories in the DS so that the beam losses with $Pb^{82+}$ ion beam collisions end up on these new collimators rather than on the downstream superconducting magnets. In IR1/5, losses are steered towards the connection cryostat of the DS that does not contain active magnetic elements, so no DS collimator is needed here. The installation of the new DS collimators relies on the development of new connection cryostats that allow the insertion of collimation equipment at room temperature.
Two of these devices have been installed during LS2 next to the ALICE experiment. Figure~\ref{fig:DSCollimator} shows a picture of this new collimator assembly during installation. For the cleaning insertion in IR7, the losses in the dispersion suppressor will be mitigated for $Pb^{82+}$ operation via new Crystal collimators that enhance the cleaning efficiency in IR7 for ion beam operation. Figure~\ref{fig:Goniometers} shows the new goniometers and the crystal collimators during their installation in the LHC.

Recent beam studies in the LHC indicate that additional DS collimators in IR7 are not required for magnet protection during nominal proton beam operation. The current setup with crystal collimators in IR7 therefore has proven to be an adequate upgrade path for the moment, confirming the decision to descope the installation of 11 T dipole magnets and dedicated DS collimators in IR7.

\subsubsection{New Superconducting Link}
The power converters and electronic racks for the magnet powering and protection of all new HL-LHC magnets will be installed in the new underground galleries, approximately 70 m to 100 m away from the~actual magnets in the tunnel. The connection between the new power converters and the new magnets in the LHC tunnel will be established with novel superconducting links that utilize high temperature superconductors based on MgB2 technology. The MgB2 superconductor is cooled by gaseous He that is evaporated inside the cryo lines of the new insertion magnets, providing temperatures between 25 K and 50 K inside the flexible cryostat of the superconducting link. The connections between the superconducting link to the power converters in the new galleries and the magnet cryostats in the tunnel are provided by new feed boxes that utilize as well novel high temperature superconductors for these transitions. 
The HL-LHC is therefore not only a performance upgrade for the LHC, but also a seedbed for various essential accelerator technologies that will find applications well beyond that of the HL-LHC project.

\begin{figure}[ht]
\begin{center}
\includegraphics[width=8.6cm]{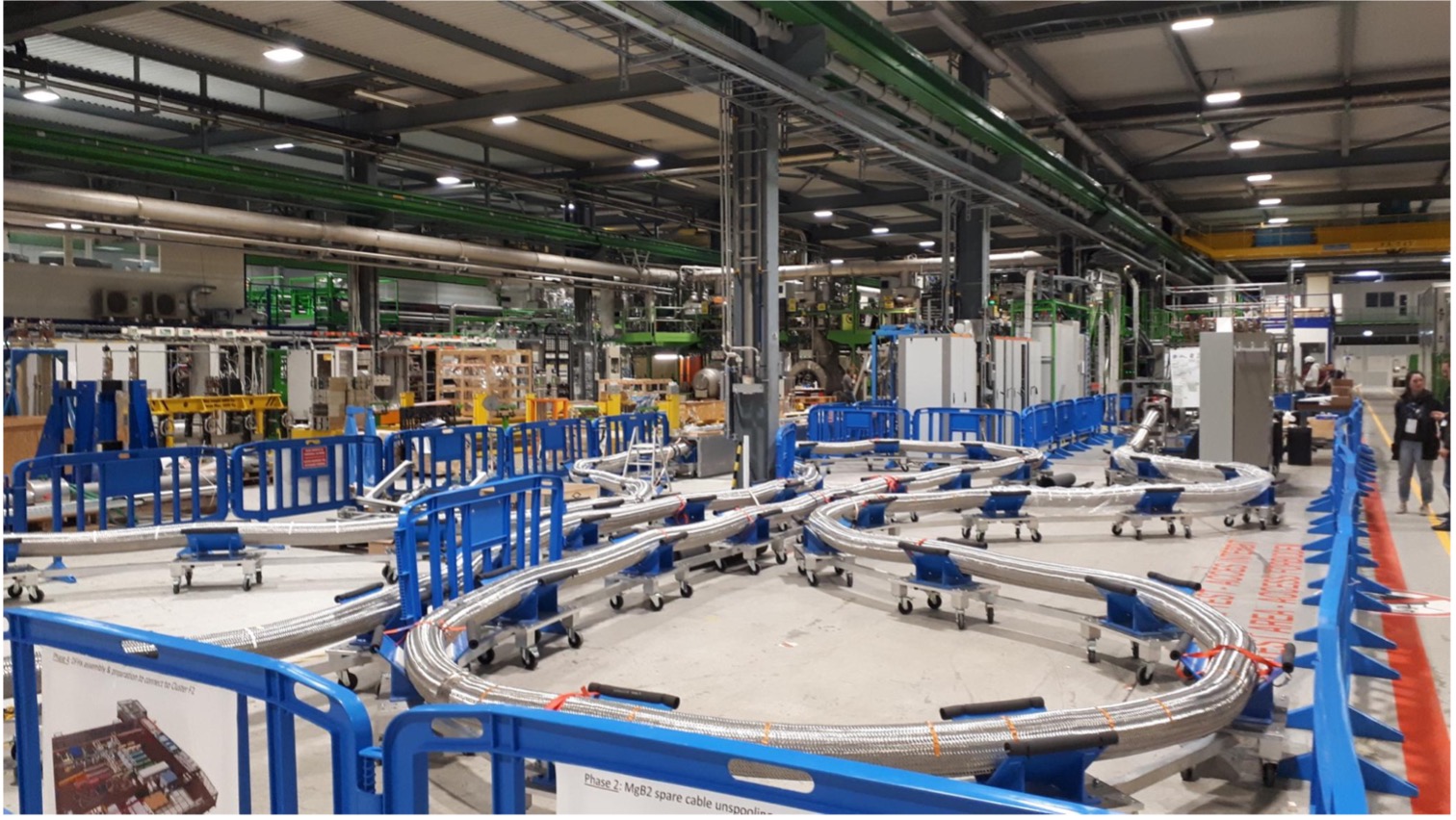} \includegraphics[width=6.5cm]{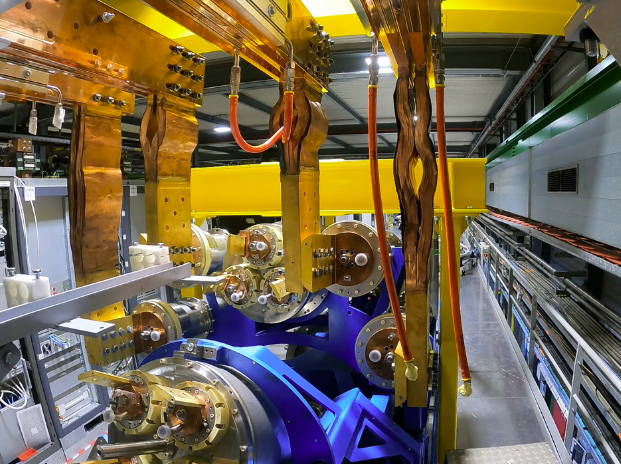}
\caption{Superconducting Link on the test bench in SM18 (left) and installed on the IT String platform (view of the connection cryostat DFHX (right).}
\label{fig:SCLink}
\end{center}
\end{figure}

Prototypes of these links have been tested at the end of 2020 [Demo-2] and beginning of 2023 [Demo-3], demonstrating the ability to transport over 100 kA of continuous currents without Ohmic losses at a temperature between 25 K and 50 K over a distance of more than 70 meters. Figure~\ref{fig:SCLink} shows the prototype system being prepared along with the DFHX prototype for testing on the F2 Testbench in SM18 at CERN and after installation in the IT String. The assembly includes the assembly of novel distribution feed boxes, which are housing the high-temperature superconducting current leads that ensure the transition to the room temperature environment and the connection of the warm powering equipment. The project is now in the phase of series production for all components of the superconducting link, and the cryostating of the first series system took place in December 2024.

\section{Summary}
The LHC machine has achieved remarkable performance levels, exceeding the nominal performance estimates in many areas by significant factors and delivering close to 1.5 $fb^{-1}$ per day during its best performance periods. Today, the LHC is routinely operating with instantaneous luminosities more than twice its design value. The accumulated radiation damage in the focusing elements next to the high luminosity experiments implies however a replacement of the existing quadrupole magnets at the end of the nominal LHC running period, that is currently projected for the mid of 2026. The HL-LHC project developed suitable new quadrupole magnets with a larger radiation tolerance compatible with a tenfold increase in the total integrated luminosity of the LHC, from approximately 500 $fb^{-1}$ from the LHC operation to up to 4000 $fb^{-1}$ [the nominal performance target for the HL-LHC machine is the production of 3000 $fb^{-1}$, but the machine hardware has been designed to be compatible with the production of 4000~$fb^{-1}$] and a suite of new technologies to boost the LHC performance to the required performance levels for achieving this integrated luminosity by the beginning of the 2040ies \cite {Operation}. Table~\ref{tab:LHC-HL-Param} compares some key machine parameters of the LHC with the upgraded HL-LHC machine.

\begin{table}[H]
\begin{center}
\caption{LHC and HL-LHC key machine parameters.}
\label{tab:LHC-HL-Param}
\begin{tabular}{p{5cm}cccc}
\hline\hline
\textbf{Parameter}& \textbf{Nominal LHC}
                                    & \textbf{LHC Achieved} 
                                         & \textbf{HL-LHC} 
                                                & \textbf{HL-LHC Ultimate}\\
\hline
Beam Energy in Collision [TeV]  & 7   & 6.8  & 7 & 7.5 \\
Number of particles per bunch   & 1.15 $10^{11}$    & 1.8 $10^{11}$  & 2.2 $10^{11}$ & 2.2 $10^{11}$\\
Number of bunches     & 2808    & 2556  & 2748 & 2748\\
Beam Current [A]       & 0.58    & 0.71  & 1.1 & 1.1\\    
Revolution frequency [kHz]      & 11.245   & 11.245 & 11.245 & 11.245 \\
$\beta$* [m]     & 0.55    & 0.3  & 0.15 & 0.10\\
Full Crossing Angle [rad]  & 285    & 320  & 500 & 500\\
Transverse Emittance [\textmu m]  & 3.75    & 2.5  & 2.5 & 2.5\\
Stored Beam Energy [MJ]  & 362    & 425  & 677 & 725\\
Events per bunch crossing  & 27    & 60  & 140 & 200 \\
Luminosity [$cm^{-2} s^{-1}$]  & 1.0 $10^{34}$    & 2.2 $10^{34}$  & 5.0 $10^{34}$ & 7.5 $10^{34}$\\
\hline\hline
\end{tabular}
\end{center}
\end{table}

\section*{Acknowledgements}
The authors wish to thank all colleagues at CERN and across the international collaborations that contribute to the realisation of the HL-LHC project.

\end{document}